\documentclass[fleqn,usenatbib]{mnras}

\usepackage{newtxtext,newtxmath}
\usepackage[T1]{fontenc}

\DeclareRobustCommand{\VAN}[3]{#2}
\let\VANthebibliography\thebibliography
\def\thebibliography{\DeclareRobustCommand{\VAN}[3]{##3}\VANthebibliography}

\usepackage{graphicx}	% Including figure files
\usepackage{amsmath}	% Advanced maths commands

\usepackage{amssymb}	% Extra maths symbols
\usepackage{xcolor}
\definecolor{green}{rgb}{0.0, 0.5, 0.0}

\title[AGN Jets with different comsic-ray composition]{Evolution and feedback of AGN Jets of different Cosmic-ray Composition}

\author[Lin et al.]{
Yen-Hsing Lin$^{1,2}$\thanks{E-mail: julius52700@gmail.com},
H.-Y. Karen Yang$^{2,3}\thanks{E-mail: hyang@phys.nthu.edu.tw}$,
Ellis R. Owen$^{4,2,5}$
\\
$^{1}$Interdisciplinary Program of Science (IPS), National Tsing Hua University, No.\ 101, Section 2, Kuang-Fu Road, Hsinchu, 30013, Taiwan\\
$^{2}$Institute of Astronomy, National Tsing Hua University, Hsinchu 30013, Taiwan\\
$^{3}$Physics Division, National Center for Theoretical Sciences, Taipei 106017, Taiwan\\
$^{4}$Theoretical Astrophysics, Department of Earth and Space Science, Graduate School of Science, Osaka University, Toyonaka,
Osaka 560-0043, Japan \\
$^{5}$Center for Informatics and Computation in Astronomy, National Tsing Hua University, Hsinchu 30013, Taiwan
}

\date{Accepted XXX. Received YYY; in original form ZZZ}

\pubyear{2022}

\begin{document}
\label{firstpage}
\pagerange{\pageref{firstpage}--\pageref{lastpage}}
\maketitle

\begin{abstract}
Jet feedback from active galactic nuclei (AGN) is one of the most promising mechanisms for suppressing cooling flows in cool-core clusters. However, the composition of AGN jets and bubbles remains uncertain; they could be thermally dominated, or dominated by cosmic-ray proton (CRp), cosmic-ray electron (CRe), or magnetic energy. 
%\textcolor{green}{In this work, }
In this work, we investigate the evolution and feedback effects of CRp and CRe dominated jets by conducting 3D magnetohydrodynamic simulations of AGN jet-inflated bubbles in the intracluster medium using the FLASH code. We present the evolution of their energies, dynamics and heating, and model their expected cavity-power versus radio-luminosity relation ($P_{\rm cav}-L_R$).
We find that bubbles inflated by CRe dominated jets follow a very similar dynamical evolution to CRp dominated bubbles even though CRe within bubbles suffer significantly stronger synchrotron and inverse-Compton cooling. This is because, as CRe lose their energy, the jet-inflated bubbles quickly become thermally dominated within $\sim 30$ Myr. Their total energy stops decreasing with CR energy and evolves similarly to CRp dominated bubbles.  
The ability of CRe and CRp dominated bubbles to heat the intracluster medium is also comparable; the cold gas formed via local thermal instabilities is well suppressed in both cases. The CRp and CRe bubbles follow different evolutionary trajectories on the $P_{\rm cav}-L_R$ plane, but the values are broadly consistent with observed ranges for FRI sources. We also discuss observational techniques that have potential for constraining the composition of AGN jets and bubbles.
\end{abstract}


%%%%%%%%%%%%%%%%%%%%%%%%%%%%%%%%%%%%%%%%%%%%%%%%%%

%%%%%%%%%%%%%%%%% BODY OF PAPER %%%%%%%%%%%%%%%%%%

\section{Introduction}\label{sec:Intro}

The immense energy output from active galactic nuclei (AGN) is believed to have a great impact on galaxy evolution across cosmic time. Specifically, relativistic jets greatly alter the evolution of galaxy clusters in the Universe by heating the intracluster medium (ICM) and preventing the runaway cooling of cool-core (CC) clusters via self-regulated feedback cycles \citep[][]{McNamara12, Blandford2019}. The detailed interaction between AGN jets and the ICM is likely to be dependent on the energy content of AGN jet-inflated bubbles \citep[e.g.,][]{Yang19}; however, the composition of AGN jets and bubbles remains poorly constrained observationally. 

When jets travel beyond kpc scales and punch into the ICM, they create low density "bubbles" that rise upward in the cluster potential through buoyancy. Due to their low density, these bubbles form low surface brightness regions in the X-ray emitting ICM, often called \textit{X-ray cavities}. By contrast, the relativistic particles inside the bubbles produce radio synchrotron emission when they interact with magnetic fields, forming the \textit{radio lobes}\footnote{ Following convention in the context of AGN feedback in galaxy clusters, the term "bubbles" in this paper refers to "X-ray cavities," regardless of whether they contain active or inactive radio sources.}. 
With high spatial resolution X-ray and radio imaging, one can measure the pressure of ICM (denoted as $P_{\mathrm{ext}}$ in this work) as well as the pressure provided by the synchrotron emitting electrons in the bubbles (denoted as $P_{\mathrm{int}}$). Interestingly, observations found that there are two populations of AGN bubbles \citep[][]{Dunn04,DeYoung06,Birzan08,Croston18}. One population presents similar external and internal pressures ($P_{\mathrm{int}}\sim P_{\mathrm{ext}}$), meaning that the pressure from the radio-emitting cosmic-ray electrons (CRe) is sufficient to support the bubbles. On the other hand, other AGN bubbles show substantially smaller pressure from CRe than the external ICM pressure ($P_{\mathrm{int}}\ll P_{\mathrm{ext}}$), suggesting that the dominant energy content inside these bubbles may come from other sources, such as magnetic fields, ultra-hot thermal plasma or cosmic-ray protons (CRp). 

Theoretically, comprehensive studies of kinetic-energy dominated jets have been performed using hydrodynamic (HD) simulations \citep[e.g.][]{Gaspari12, Barai14, Reynolds15, Hillel16, Yang16a, Yang16b, Barai16, Li17, Fabian17, Bambic18}. 
More recently, the properties of CRp dominated jets have also seen growing attention in the literature \citep[e.g.][]{Mathews08, Guo11, Ruszkowski17, Ehlert18, Yang19, Su2021, Beckmann22}. These works have demonstrated that differences in the energy content of AGN bubbles could lead to very different heating impacts and dynamical evolution of the ICM. For instance, it is shown that CRp dominated bubbles tend to be more oblate and buoyant, leading to more efficient uplift of the ICM and suppressed radiative cooling \citep{Guo08, Mathews08, Yang19}. Direct mixing between the ICM and ultra-hot thermal plasma could be the dominant heating mechanism in kinetic-energy dominated jets. For CRp dominated jets, the CRs could instead heat the ICM via Coulomb, hadronic collisions and streaming heating \citep[e.g.][]{Ruszkowski17, Yang19}. As a result, self-regulated AGN feedback simulations including kinetic-energy dominated jets tend to produce more quasi-steady AGN activities \citep[e.g.,][]{Yang16b}, while simulations including CRp dominated jets tend to exhibit more episodic AGN activities \citep{Ruszkowski17}. Therefore, understanding the composition of AGN jets/bubbles and their impact on the dynamics and heating of the ICM is crucial for predicting the behavior of feeding onto the central AGN and the evolution of galaxy clusters under the influence of AGN feedback.

While the feedback effects of kinetic-energy and CRp dominated jets are better understood, those of CRe dominated jets have not yet been investigated in the literature. The existence of CRe dominated bubbles can be inferred from observations of AGN bubbles that are relatively close to pressure balance without the need for non-radiating CRs ($P_{\mathrm{int}}\sim P_{\mathrm{ext}}$) \citep{Dunn04, Croston18}. For the Fanaroff-Riley \citep[FR;][]{FR1974MNRAS} I sources often found in galaxy clusters, the CRe within the bubbles/lobes can be supplied by freshly injected CRe at the shocks/flaring point when they travel to kpc scales \citep[see e.g. Section 3.1.3 in][]{Blandford2019}. However, the long-term evolution of these CRe bubbles and their impact on AGN feedback remains an open question.

It is therefore our aim to investigate the differences between CRp and CRe dominated jets and their subsequent evolution and feedback impact on the ICM. There are several reasons why one may expect them to show very different feedback behaviors. Firstly, compared to CRp, CRe suffer stronger synchrotron and inverse-Compton (IC) losses within magnetic fields and cosmic microwave background (CMB) radiation, respectively. Secondly, CRp can heat the ICM via hadronic and Coulomb collisions, while CRe cannot undergo hadronic heating and their Coulomb heating effect is comparatively negligible.

Therefore, one might expect to find an interesting case of failed AGN feedback, where CRe dominated bubbles eventually deflate, and fail to provide sufficient heating to the ICM. To this end, we perform three-dimensional (3D) magnetohydrodynamic (MHD) simulations including relevant CR physics to investigate the energy, dynamics and heating of CRe bubbles and compare their evolution with their CRp counterparts.

The paper is organized as follows. In Section \ref{sec:Method}, we describe the governing equations, assumptions, initial conditions, and the detailed modeling for CRp and CRe. In Section \ref{sec:Result}, we present the overall evolution (Section \ref{sec: Bubble_Evo}) of the bubbles, the evolution of different energy components in the bubbles (Section \ref{sec: Energy_Evo}), and heating/cooling profiles and cold gas formation (Section \ref{sec:heating_coldgas}). We compare the radio-luminosity versus cavity-power relations predicted by our simulations with observed data in Section \ref{sec:LRQjet}. In Section \ref{sec:Discussion}, we discuss the implications of our results to the current understanding of jet composition, followed by the conclusions in Section \ref{sec:Conclusion}.

\section{Methods}\label{sec:Method}
We perform 3D MHD simulations of AGN jet-inflated bubbles in an idealized Perseus-like cluster using the FLASH code \citep{Flash, Dubey08}. We focus on two cases: CRp dominated jets and CRe dominated jets (hereafter CRp and CRe jets, respectively). The main difference between the two is the cooling of CRs and the amount of heating they provide to the thermal gas.
We also investigate the effects of CR streaming by performing two additional simulations (called CRpS and CReS, respectively). Detailed physics included in each simulation is summarised in Table \ref{tab:Sim_Type}.

\begin{table}
    \begin{tabular}{|c|c|c|c|}
         \hline
         Simulation name & Streaming & IC + sync.\ cooling & Coulomb and \\
         & & & hadronic cooling\\
         \hline
         CRp & No & No & Yes\\
         CRe & No & Yes & No \\
         CRpS & Yes & No & Yes\\
         CReS & Yes & Yes & No\\
         \hline
    \end{tabular}
    \centering
    \caption{Summary of the 4 main simulations we performed and the physics adopted in each of them.}
    \label{tab:Sim_Type}
\end{table}

\subsection{Cosmic-ray physics}

We treat CRs as a second fluid that follows the MHD equations \citep[see e.g.][for more detailed descriptions]{Yang12b, Yang19}:

\begin{equation}
\frac{\partial \rho}{\partial t}+\nabla \cdot(\rho \boldsymbol{v})=0
\end{equation}

\begin{equation}
\frac{\partial \rho \boldsymbol{v}}{\partial t}+\nabla \cdot\left(\rho \boldsymbol{v} \boldsymbol{v}-\frac{\boldsymbol{B} \boldsymbol{B}}{4 \pi}\right)+\nabla p_{\mathrm{tot}}=\rho \boldsymbol{g}
\end{equation}

\begin{equation}
\frac{\partial \boldsymbol{B}}{\partial t}-\nabla \times(\boldsymbol{v} \times \boldsymbol{B})=0
\end{equation}

\begin{multline}
\frac{\partial e}{\partial t}+\nabla \cdot\left[\left(e+p_{\mathrm{tot}}\right) \boldsymbol{v}-\frac{\boldsymbol{B}(\boldsymbol{B} \cdot \boldsymbol{v})}{4 \pi}\right]\\
=\rho \boldsymbol{v} \cdot \boldsymbol{g}+\nabla \cdot\left(\boldsymbol{\kappa} \cdot \nabla e_{\mathrm{cr}}\right) + \mathcal{H}_{\mathrm{cr}} - n_e^2\Lambda(T)
\end{multline}

\begin{equation}
\frac{\partial e_{\mathrm{cr}}}{\partial t}+\nabla \cdot\left(e_{\mathrm{cr}} \boldsymbol{v}\right)=-p_{\mathrm{cr}} \nabla \cdot \boldsymbol{v}+\nabla \cdot\left(\boldsymbol{\kappa} \cdot \nabla e_{\mathrm{cr}}\right) + \mathcal{C}_{\mathrm{cr}}
\end{equation}
where $\rho$ and $v$ are the density and velocity of gas, respectively, $\boldsymbol{g}$ is the gravitational field, $\boldsymbol{\kappa}$ is the CR diffusion tensor, $e_{\textrm{cr}}$ is the CR energy density, and $e$ is the total energy density, consisting of kinetic, thermal, CR and magnetic energy ($e=0.5\rho v^2 + e_{\textrm{th}} + e_{\textrm{cr}} + B^2/8\pi$). The total pressure is $p_{\textrm{tot}} = (\gamma-1)e_{\textrm{th}} + (\gamma_{\textrm{cr}}-1)e_{\textrm{cr}} + B^2/8\pi$, in which we adopt the adiabatic index $\gamma=5/3$ for the thermal gas and $\gamma_{\textrm{cr}}=4/3$ for relativistic CRs.
$\mathcal{H}_{\mathrm{cr}}$ is the net contribution of CR related processes to the change of total energy density, $\mathcal{C}_{\mathrm{cr}}$ is the CR cooling rate due to the combined effect of Coulomb losses, hadronic processes, streaming, IC scattering and synchrotron losses. $n_e$ is the electron number density, and $\Lambda(T)$ is the radiative cooling function.

In the above CR-MHD formalism \citep{Zweibel13, Zweibel17}, to the first order, CRs advect with the thermal gas since they are well scattered by small-scale structures in the magnetic field. Assuming the CRs are primarily scattered by waves as a part of a background turbulent magnetic field and that the turbulence is isotropic, there is no net energy transfer from the CRs to the gas. This picture, called the \textit{extrinsic turbulence} picture of CR transport, is adopted in many early studies of CR simulations \citep[e.g.][]{Guo08, Mathews08}. 
In this picture, one can approximate CR transport as a spatial diffusion process, with a diffusion coefficient of $\kappa \sim 3\times 10^{28} {\rm cm ~s^{-1}}$, which is typical for an ICM environment \citep{Yang19}. 
This model (identical to the CRdh simulation in \citealt{Yang19}) is adopted in the CRp and CRe simulations as listed in Table \ref{tab:Sim_Type}.

For the CRpS and CReS simulations, we adopt the \textit{self-confinement} picture of CR transport. In this picture, CRs are assumed to be scattered by their self-excited Alfvén waves via the streaming instability \citep{Kulsrud69, Wentzel74, Zweibel13}. Direct simulations of streaming have been performed in previous works \citep[e.g.][]{Ruszkowski17, Jiang2018ApJ, Thomas2019MNRAS} but they are more numerically challenging. Nevertheless, under the assumption that CRs are well scattered by small-scale magnetic field structures that are unresolved, one can also approximate CR transport due to streaming as spatial diffusion, and one can show that the CR diffusion coefficient is comparable to that in the \textit{extrinsic turbulence} picture \citep[see][for more detailed discussion]{Yang19}. For simplicity, this approximation is applied in our CRpS and CReS simulations. In addition to the CR transport term, streaming can also transfer energy from the CRs to Alfvén waves and subsequently heat up the thermal gas. This corresponds to the CR cooling term in the CR energy density equation, $\mathcal{C}_{\mathrm{cr, s}}=\boldsymbol{v}_A \cdot \nabla p_{\mathrm{cr}}$. Note that streaming transfers energy from the CRs to the thermal gas, and therefore the total energy density would be unchanged during the process (i.e., the contribution to $\mathcal{H}_{\mathrm{cr}}$ due to streaming is zero).

Following \citet{YoastHull13, Ruszkowski17}, the energy loss rates of CRp due to Coulomb and hadronic processes can be written as:
\begin{equation}\label{Eq:Coulomb_Cooling}
\mathcal{C}_{\mathrm{CRp}, \mathrm{c}}=-4.93 \times 10^{-19} \frac{n-4}{n-3} \frac{e_{\mathrm{cr}} \rho}{E_{\rm min, GeV}} \frac{\rho}{\mu_{\mathrm{e}} m_{\mathrm{p}}} \operatorname{erg} \mathrm{cm}^{-3} \mathrm{~s}^{-1},
\end{equation}
and
\begin{equation}\label{Eq:Hadronic_Cooling}
\mathcal{C}_{\mathrm{CRp}, \mathrm{h}}=-8.56 \times 10^{-19} \frac{n-4}{n-3} \frac{e_{\mathrm{cr}} \rho}{E_{\rm min, GeV}} \frac{\rho}{\mu_{\mathrm{p}} m_{\mathrm{p}}} \operatorname{erg} \mathrm{cm}^{-3} \mathrm{~s}^{-1},
\end{equation}
where $n$ is the slope of the power-law distribution function of CRp in momentum space, $E_{\mathrm{min,GeV}}$ is the minimum energy of CRp in units of GeV, and $\mu_p$ and $\mu_e$ are the mean molecular weights per proton and electron, respectively. 

In hadronic processes, secondary electrons receive around $1/6$ of the inelastic energy which is transferred to heat the gas \citep{Mannheim94, Guo08}, see also \citet{Owen2018MNRAS}, which considered the efficiency of this heating in different conditions; the remainder will be emitted and lost to gamma rays or neutrinos via pion production processes. Thus,  $\mathcal{H}_{\mathrm{CRp}}=(5/6)\;\!\mathcal {C}_{\mathrm{CRp}, \mathrm{h}}$.

Following \citet[][]{YoastHull13,Miniati2001}, the cooling rate of CRe due to synchrotron emission and IC scattering can be written as
\begin{equation}\label{eq:C_cr}
\mathcal{C}_{\mathrm{CRe}, \mathrm{IC+Syn}}=\frac{2-p}{3-p}\beta e_{\mathrm{cr}}\left[ \left( \frac{E_{\textrm{max}}}{E_{\textrm{min}}} \right)^{2-p} -1 \right]\left( E_{\textrm{max}} - E_{\textrm{min}} \right),
\end{equation} 
where it is assumed that the energy spectrum of CRe follows a power-law distribution, and $\beta= 4\sigma_\mathrm{T} (U_\mathrm{B} + U_\mathrm{r})/(3m_e^2 c^3)$, in which $\sigma_\mathrm{T}$ is the Thomson scattering cross section, $U_\mathrm{B}= 4\times 10^{-14}(B/\mu {\rm G})^2{\rm ~erg~cm^{-3}}$ is the magnetic energy density ($B$ is the magnetic field strength), and $U_\mathrm{r} = 4.2\times 10^{-13}(1+z)^4~\mathrm{erg~cm^{-3}}$ is the energy density of the CMB. We set $z=0$ for all our simulations since the observed cavity systems of interest are primarily at low redshift.
$p=2.5$ is the slope of CRe distribution function ($n(E)\propto E^{-p}$), and $E_{\textrm{max}}$ and $E_{\textrm{min}}$ are the maximum and minimum energy of the CRs, respectively. 

In general, $\mathcal{C}_{\mathrm{CRe}, \mathrm{IC+Syn}}$ at any given simulation time step depends on the local CRe spectrum ($E_{\rm max}$ and $E_{\rm min}$) and the magnetic field strength. While in our MHD simulations, the magnetic field is self-consistently modeled, the values of $E_{\rm max}$ and $E_{\rm min}$ for the CRe need to be assumed. Strictly, one would need to follow the evolution of the CRe spectrum on-the-fly during the simulations in order to compute the CRe energy losses self-consistently \citep[e.g.][]{Yang17}. However, the system we are modeling is relatively simple, where the CRe spectral evolution is dominated by synchrotron and IC cooling after the initial jet injection phase. As such, adopt a simpler approach to model $E_{\rm max}$ and $E_{\rm min}$. Specifically, we assume that the energy spectrum of CRe starts with a power-law distribution with $E_{\textrm{max},0}=100$ GeV and $E_{\textrm{min},0}=1$ GeV from $t=0$ Myr up to $t=10$ Myr (the inflation period). Then, we evolve $E_{\textrm{max}}$ and $E_{\textrm{min}}$ according to 
\begin{equation}
    E = \frac{E_0}{1 + \beta( t-10~\mathrm{Myr}) E_0},
    \label{eq:E_lim_evo}
\end{equation} 
\citep{Kardashev1962}, 
where $\beta$ is the same as described by Eq. \ref{eq:C_cr}, and $t$ is the simulation time. We offset the start time of the spectral evolution to 10 Myr because, during the injection phase, there can be a complex mix of processes (e.g., mixing between newly injected CRe and existing CRe, adiabatic expansion, and possibly in situ acceleration) that are not well constrained. These uncertainties can be effectively captured by using different values of $E_{\rm max,0}$ and $E_{\rm min,0}$ (see Appendix \ref{sec:EmaxEmin}). Equation~\ref{eq:E_lim_evo} is only valid for a constant magnetic field strength. Therefore, for consistency, we adopt a field strength of $1~ \mu$G to model the evolution of $E_{\rm max}$ and $E_{\rm min}$, instead of using the magnetic fields from our simulations. We verified this approach to be reasonable by comparing results using simulated magnetic field strengths and constant field strengths. We found these are not significantly different since IC cooling is much stronger than synchrotron cooling in our setup.

\subsection{Simulation setup}
We perform the simulations in a box of 500 kpc on each side. The simulation domain is refined adaptively up to 8 levels of refinement (corresponding to a maximum resolution of 500 pc), according to a refinement criterion based on steep temperature gradients. The total simulation time is 100 Myr. The initial gas profile of the cluster is set using empirical fits with the Perseus cluster, one of the most studied CC clusters in the local Universe, in hydrostatic equilibrium within a static Navarro-Frenk-White \citep[NFW, ][]{Navarro96} gravitational potential. Radiative cooling of the gas is calculated using tables from \citet{SutherlandDopita}, adopting $1/3$ solar metallicity. The initial cooling time of the simulated Perseus-like cluster is around 250 Myr. Although this radiative cooling time is longer than the simulation time of 100 Myr, we note that it is still essential to include radiative cooling. This is because it would affect the amount of cold gas formed via local thermal instabilities \citep[e.g.,][]{McCourt11}.
A reflective boundary condition is chosen for the simulations, so that the total energy within the simulation domain after the initial jet injection phase would be conserved in the absence of cooling, or otherwise lost only due to radiative cooling as in our current study\footnote{Our conclusions are insensitive to the choice of the boundary condition since the region of interest is much smaller than the simulation box size.}.
We generate a tangled ICM magnetic field by conducting a 3D inverse Fourier transform of a magnetic power spectrum with coherence length of 50 kpc \citep[see e.g.][for detailed descriptions]{Yang17}. We then normalize the field strength to the local thermal pressure so that the plasma beta is constant, $\beta=P_{\mathrm{th}}/P_\mathrm{B}=100$ \citep{Carilli02}. 

Following the method of AGN jet injection in \cite{Yang19}, bipolar AGN jets are injected from a cylinder with radius of 2 kpc and height of 4 kpc at the center of the simulation box. The power of each jet is $\dot{E}_{\mathrm{ej}}=5\times 10^{45}$ erg s$^{-1}$ and the duration of their activity is 10 Myr. These parameter choices are representative of those obtained from previous self-regulated feedback simulations using a similar setup \citep{Yang16b}. They are also consistent with recent estimated ranges of kinetic luminosities and duty cycles of a large sample of radio galaxies \citep{Shabala2020MNRAS,Hardcastle2019A&A}. The rate of mass injection can be written as $\dot{M}_{\mathrm{ej}}=2(1-f_{\mathrm{cr}})\dot{E}_{\mathrm{ej}}/v_{\mathrm{ej}}^2$, where $v_{\mathrm{ej}}=0.01c$ is the bulk velocity of the jet, which is chosen to represent jets which have gone through significant deceleration on kpc scales \citep{Laing06}, and $f_{\mathrm{cr}}=e_{\mathrm{cr}}/e=0.9$ is the fraction of CR energy in the jets. The remaining 10 per cent of the injected energy is in kinetic form; no magnetic or thermal energy is explicitly injected.

\section{Results}\label{sec:Result}
In this section, we describe the important features in our simulations. In Section \ref{sec: Bubble_Evo}, we describe the general evolution of the bubbles. In Section \ref{sec: Energy_Evo}, we describe the evolution of different energy components in the bubbles. 
In Section \ref{sec:heating_coldgas}, we describe the radial CR heating profile in the cluster and the evolution of cold ($T\leq5\times 10^5$ K) gas formation. Finally in Section \ref{sec:LRQjet}, we compare the predicted synchrotron luminosity ($L_\mathrm{R}$) and the jet power ($P_{\mathrm{cav}}$) with observations.

\subsection{Bubble evolution} \label{sec: Bubble_Evo}
\begin{figure*}
    \centering
    \includegraphics[width=\textwidth]{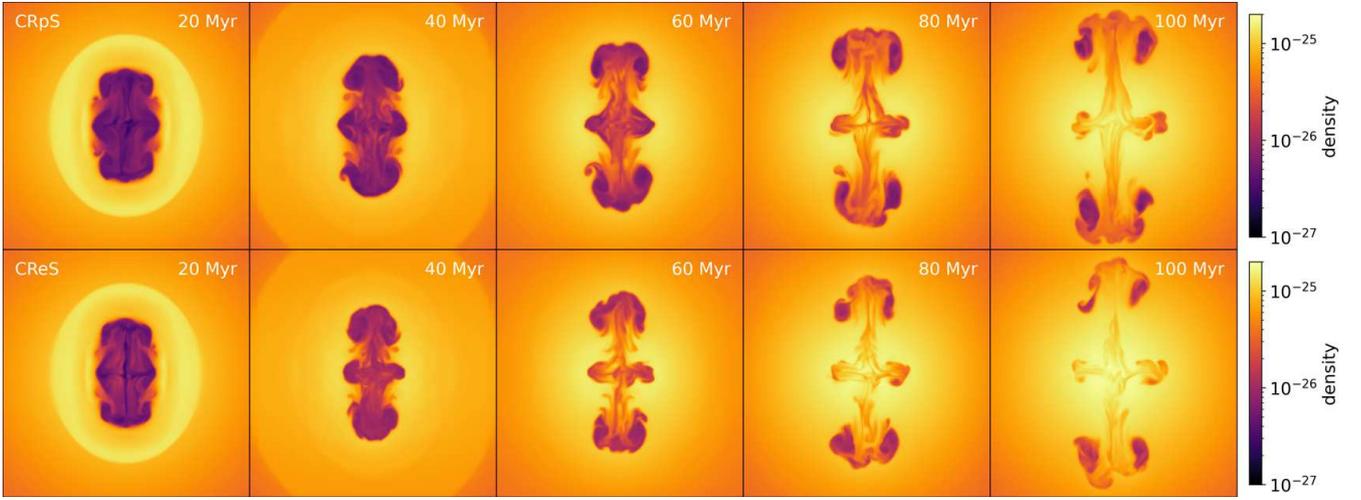}
    \caption{Density slices of the CRpS (top) and CReS (bottom) simulations at $t=20, 40, 60, 80, 100$ Myr. The physical scale of the plotted region is 100 kpc by 100 kpc.}
    \label{fig:density_evo}
\end{figure*}

Fig. \ref{fig:density_evo} shows the evolution of two representative simulations, CRpS and CReS. In their early stages, the ram pressure of the jets inflates low-density bubbles and create a series of shock waves. After the jets are turned off at $t=10$ Myr, the rapid inflation of the bubbles stops and they then rise due to buoyancy. Next, hydrodynamic instabilities (e.g. Rayleigh-Taylor and Kelvin-Helmholtz) start to deform and disrupt the bubbles due to the large shear velocity and density contrast between the bubbles and ambient medium. Eventually, the bubbles break into several small, irregular bubbles and mix with the ambient thermal gas. Contrary to our expectation, the evolution of the CRe/CReS bubbles is very similar to the CRp/CRpS bubbles. The dynamical evolution of the bubbles seems to be unaffected by the cooling differences between the two types of CRs. We discuss this result further in Section \ref{sec:Discussion}.

\begin{figure*}
    \centering
    \includegraphics[width=\textwidth]{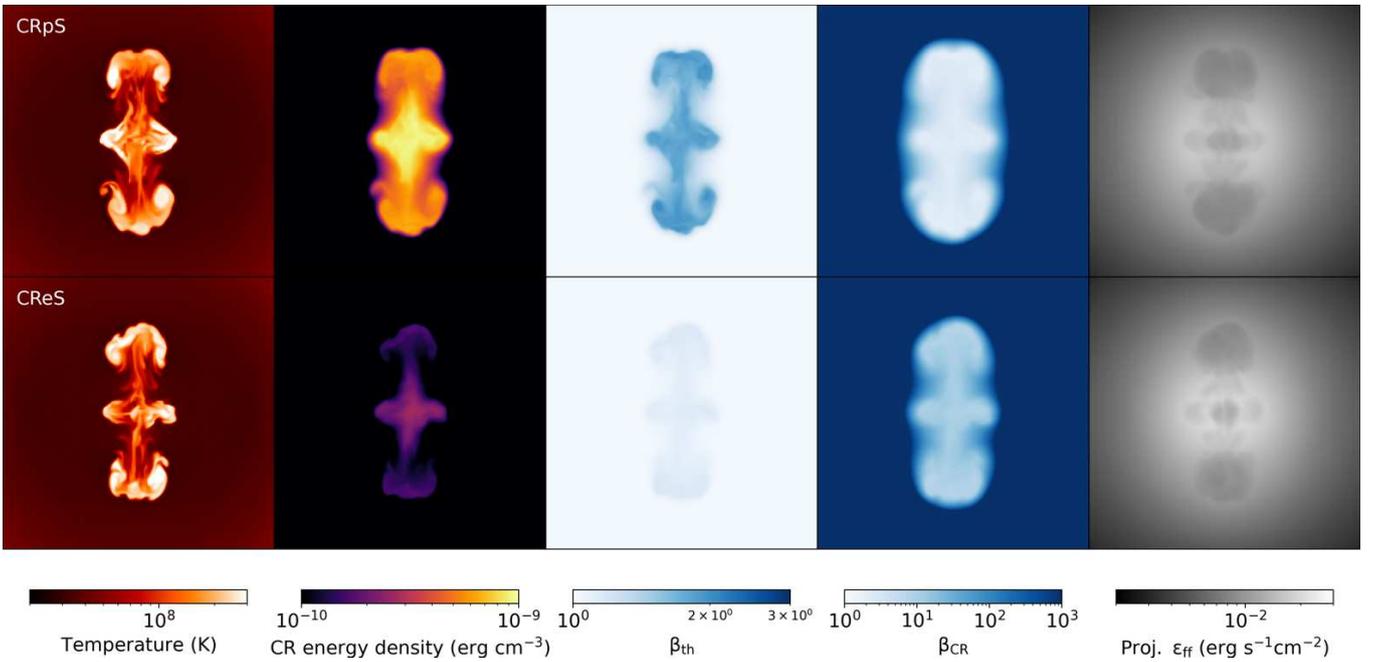}
    \caption{Different fields in the CRpS (top) and CReS (bottom) simulations at $t=60$ Myr, including temperature, CR energy density, $\beta_{\mathrm{th}}=P_{\mathrm{tot}}/P_{\mathrm{th}}$, $\beta_{\mathrm{CR}}=P_{\mathrm{tot}}/P_{\mathrm{CR}}$, and projected free-free emissivity. The plotted region is 100 kpc by 100 kpc.}
    \label{fig:fields}
\end{figure*}

In Fig. \ref{fig:fields}, we show (from left to right) slices of temperature, CR energy density, total-to-thermal pressure ratio $\beta_{\mathrm{th}}$, total-to-CR pressure ratio $\beta_{\mathrm{CR}}$, and the projected thermal Bremsstrahlung emissivity of the CRpS and CReS simulations at $t=60$ Myr.
At this epoch, the CR energy density within the CRpS bubbles is still around $10^{-9}$ erg/cm$^3$, while that of the CReS bubbles is more than 3 times weaker due to the strong synchrotron plus IC cooling of CRe. Similarly, CRpS bubbles have $\beta_{\mathrm{th}}$ around 2 to 3 and $\beta_{\mathrm{CR}}$ around 1, which means that the pressure and dynamics inside the bubbles is dominated by CR pressure. On the other hand, the CReS bubbles are dominated by thermal pressure with $\beta_{\rm th} \sim 1$. The gas temperature of the two simulations are similar: both CRpS and CReS bubbles show a peak temperature around $2\times10^8$ K at the edges of the bubbles. Finally, we use the projected thermal Bremsstrahlung emissivity for typical ICM conditions calculated using 
\begin{equation}
    \epsilon_{\mathrm{ff}} = 3 \times 10^{-27} T^{1/2} n_e^2~{\rm erg~cm^{-3}s^{-1}}
\end{equation}
\citep{Sarazin1986}
as a proxy to emulate the expected X-ray emission maps from our simulations, and find that the X-ray cavities have a similar morphology in the two cases.

\subsection{Energy evolution}\label{sec: Energy_Evo}

\begin{figure*}
    \centering
    \includegraphics[width=0.8\textwidth]{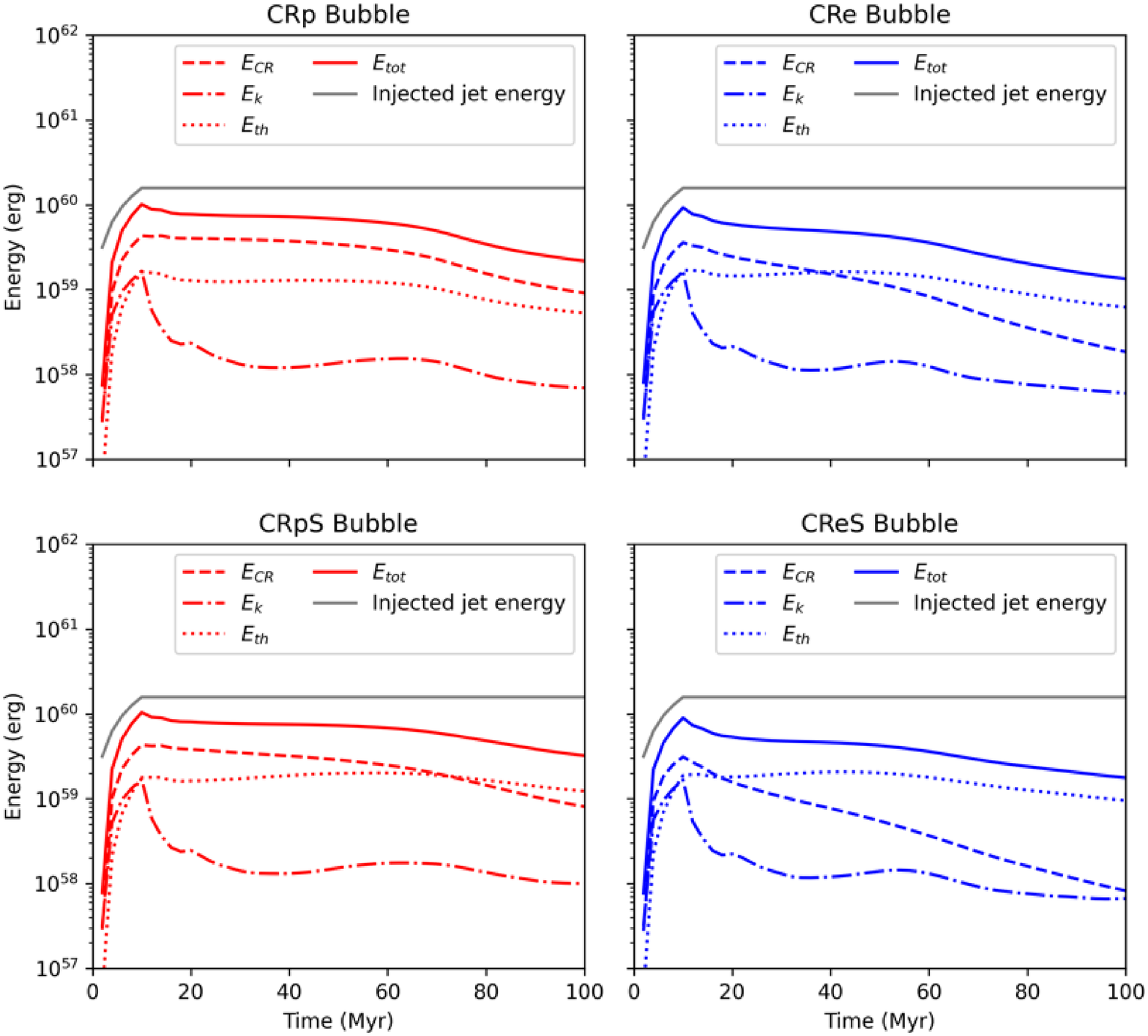}
    \caption{The evolution of different energy components inside the bubbles defined by $t_{\mathrm{cool}}\geq 3$ Gyr. The gray solid lines represent the accumulated energy injected by jets. The red lines represent CRp/CRpS, and the blue lines represent CRe/CReS. The solid, dashed, dotted, and dotted-dashed lines represent the total energy ($E_\mathrm{tot}$), CR energy ($E_\mathrm{CR}$), thermal energy ($E_\mathrm{th}$), and kinetic energy ($E_\mathrm{k}$) within the bubbles, respectively. While the CRp/CRpS bubbles are dominated by CR energy for most of their evolution, the CRe/CReS bubbles suffer more significant synchrotron and IC cooling and quickly become thermally dominated after $t\sim 30$ Myr.}
    \label{fig:E_EVO}
\end{figure*}

Fig. \ref{fig:E_EVO} shows the evolution of different energy components inside the bubbles.
We define bubbles according to the criteria $t_{\mathrm{cool}}\geq 3$ Gyr, where $t_{\rm cool}$ is the cooling timescale of the gas by Bremsstrahlung, given by 
\begin{equation}
t_{\mathrm{cool}} \sim 4.4 ~ \left( \frac{n_e}{10^{-2} \;\! {\rm cm}^{-3}} \right)^{-1} \left(\frac{T}{10^{8}\;\!{\rm K}}\right)^{1/2} \;\! {\rm Gyr} \ ,
\end{equation}
where $n_e \sim n/2 \sim \rho/m_p$. In Appendix \ref{app: BubbleDef}, we show that the choice of the $t_{\mathrm{cool}}$ threshold does not affect our main conclusions.

For the first 10 Myr, the total energy inside the bubbles rises due to the energy injection of the jets. During this period, all the bubbles are CR-energy dominant in all four simulations; the kinetic energy and thermal energy are sub-dominant. After the injection ends, the four scenarios start to differ due to their underlying physics. In the CRp simulation, the bubbles remain CR-energy dominant all the way from 10 to 100 Myr, only slightly decreasing due to the gradual expansion of the bubbles and CR energy losses via the hadronic processes. For the CRpS case, the CR energy decreases faster than the CRp case because of the additional energy transferred from the CRs to the gas via streaming. The thermal energy becomes the dominant energy component for the CRpS bubbles at around 75 Myr. By contrast, CR energy is depleted much faster in the CRe and CReS cases due to the strong synchrotron and IC cooling of CRe. The thermal and CR energies cross at around 20 to 40 Myr. In all four cases, kinetic energy remains a sub-dominant component.

\subsection{Cold gas evolution and CR heating}\label{sec:heating_coldgas}

\begin{figure*}
    \centering
    \includegraphics[width=0.8\textwidth]{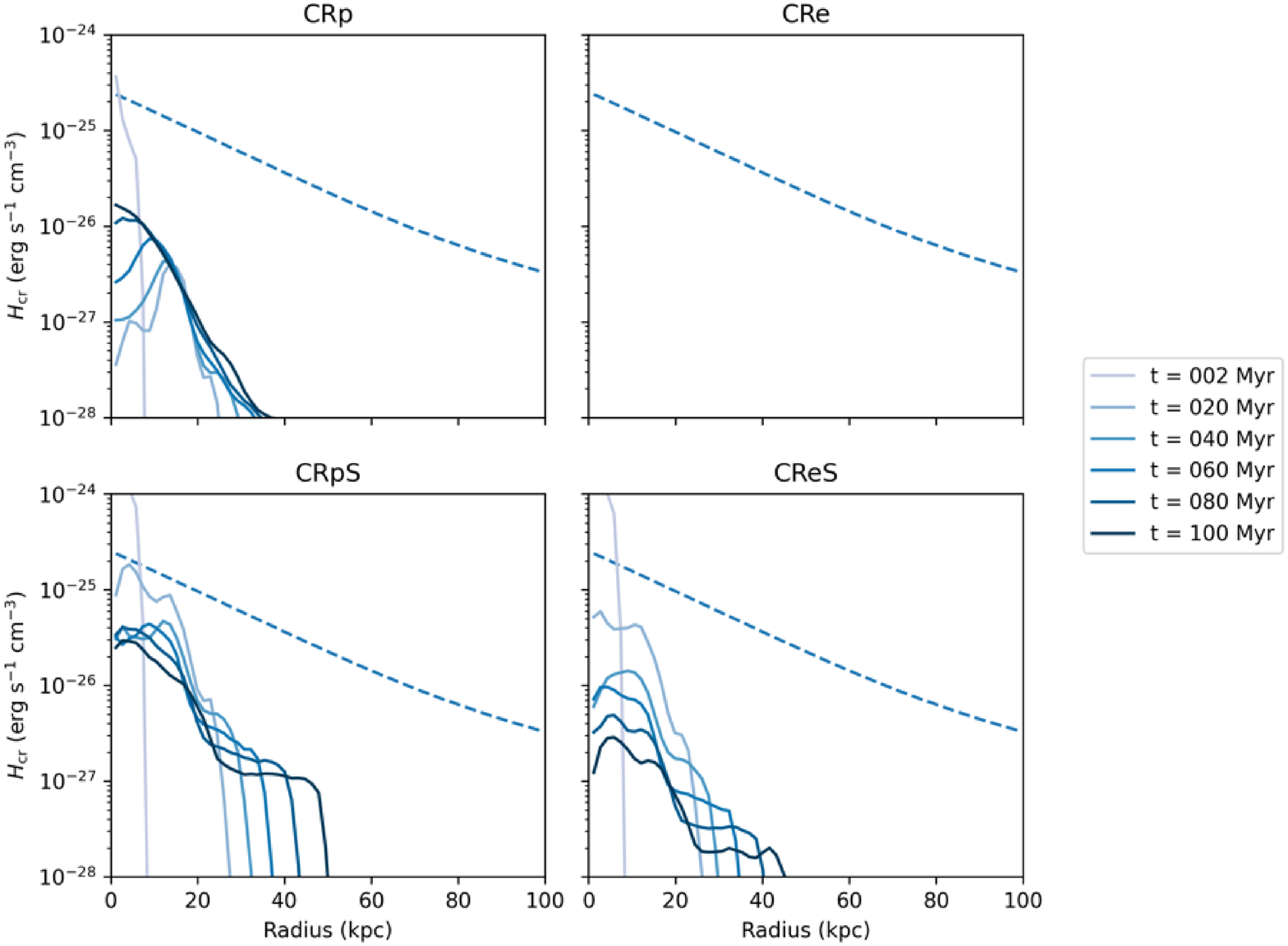}
    \caption{The radially averaged profiles of the total CR heating rates (color-coded solid lines) and the initial radiative cooling profile (dashed lines) of the four simulations at $t=2, 20, 40, 60, 80$ and 100 Myr.}
    \label{fig:heating_profile}
\end{figure*}

In Fig. \ref{fig:heating_profile}, we show the evolution of radially averaged profiles of the total CR heating rates (including hadronic, Coulomb and streaming heating) over-plotted with the radiative cooling profile of the cluster at $t=0$. In the CRe simulation, we do not consider CR heating. This is because hadronic heating would not operate, while Coulomb heating would be very inefficient for the conditions of the ICM \citep[see e.g. Fig. 4 of][]{Owen2018MNRAS}.
In the CRp simulation, the CR heating profile decreases, then subsequently increases. This behavior originates from the rapid drop of density due to bubble inflation at the cluster center in the early epoch. After the bubbles form and detach, the CRs gradually diffuse out and interact with the ambient ICM. This increases the hadronic and Coulomb heating rates. For the CRpS case, the CR heating profiles are somewhat different to the CRp case because streaming heating dominates over hadronic and Coulomb heating \citep{Ruszkowski17}. Overall the heating rates in the CRpS case are greater than the case without streaming, and the heated region extends to larger radii due to additional streaming transport. CReS bubbles behave similarly to the CRpS bubbles, but the heating is weaker. This is because cooling removes most of the CRe energy. 

\begin{figure}
    \centering
    \includegraphics[width=\columnwidth]{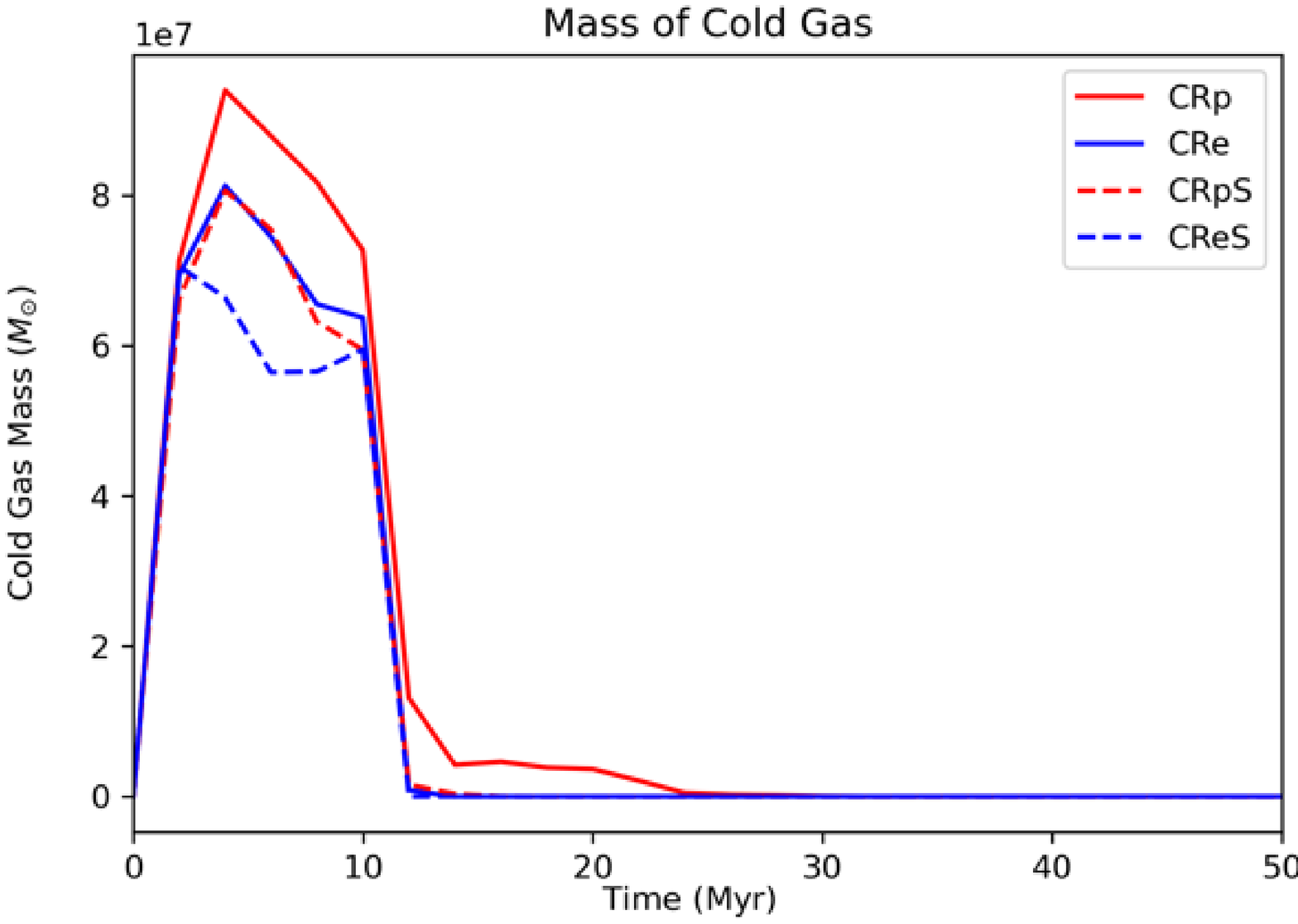}
    \caption{The evolution of cold-gas mass in the simulations. CRp/CRpS simulations tend to create rougly 20\% more cold gas. Including streaming heating can help suppress cold-gas formation by around 20\%. Values after 50 Myr are zero in all simulations.}
    \label{fig:coldmass5e5}
\end{figure}

In all four simulations, the CR heating rates are much smaller than the radiative cooling rate. Nonetheless, as shown in Fig. \ref{fig:coldmass5e5}, cold ($T\leq 5\times 10^5$ K) gas formed during the jet-injection phase via thermal instabilities triggered by rapid adiabatic cooling is well suppressed in all cases. Cold gas only forms in the first 12 Myr in the simulations due to the strong adiabatic cooling at early stages and is subsequently heated through direct mixing between the hot bubbles and the ambient gas. This suggests that even with streaming heating included, direct mixing is still the dominant heating mechanism \citep{Hillel16, Yang16b}. This conclusion is further supported by the CRe case, in which there is no CR heating, and the heating can only come from direct mixing. Overall, streaming only helps to suppress the amount of cold gas mass below $T\le 5\times 10^5$K by $\sim 20$ per cent. 

\subsection{Observable properties in radio}\label{sec:LRQjet}

\begin{figure*}
    \centering
    \includegraphics[width=0.8\textwidth]{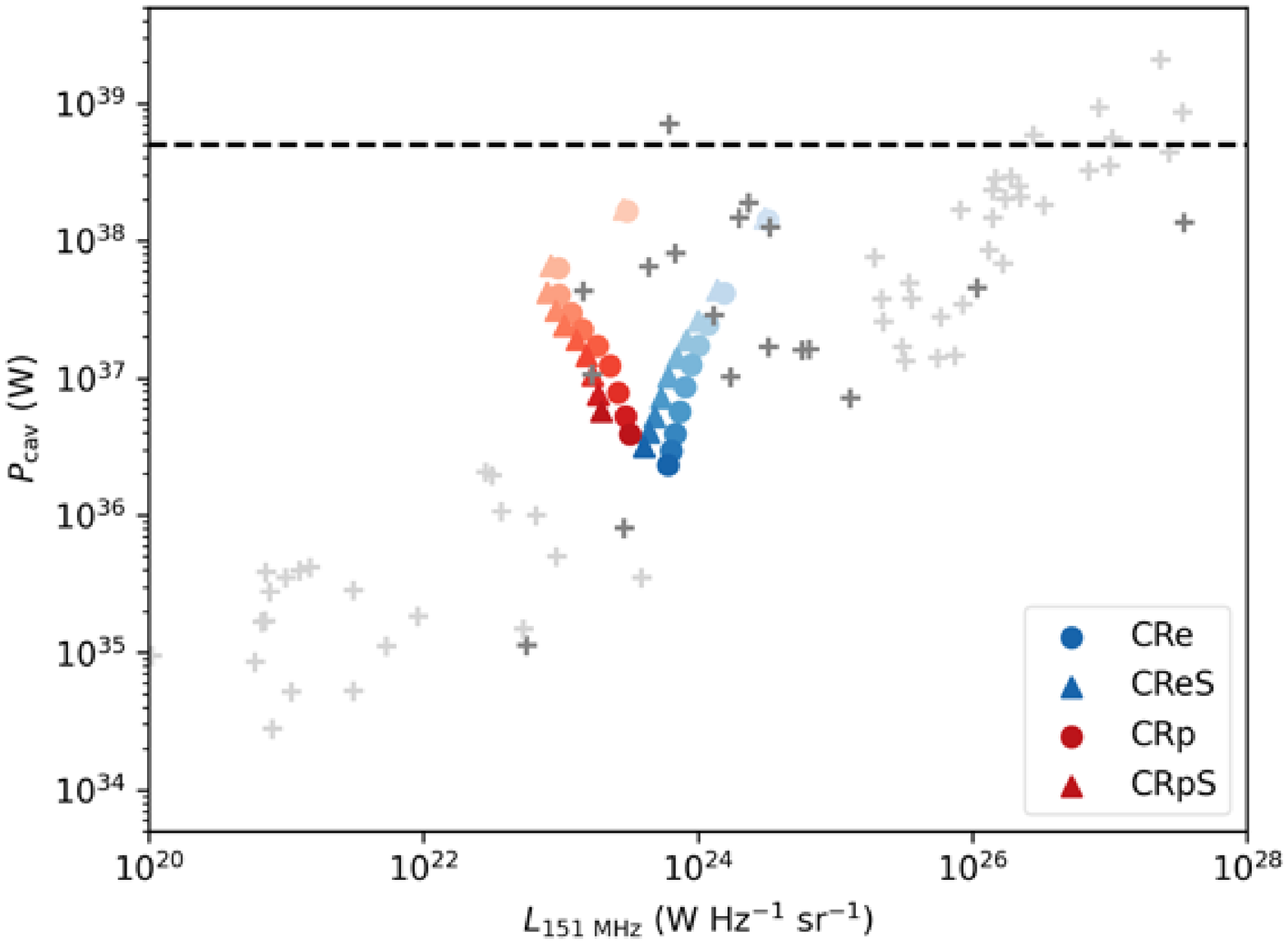}
    \caption{$P_{\mathrm{cav}}-L_{\mathrm{R}}$ diagram of our simulated bubbles over-plotted with observational data from \citet{Croston18}. The data points obtained from our simulations are plotted from $t=$ 5 to 95 Myr with a time interval of 10 Myr (where lighter colors represents earlier simulation times). We mark the observational data from \citet{Birzan08} using dark gray crosses while other data \citep[from][]{Cavagnolo2010ApJ,O'Sullivan2011ApJ, Ineson2017MNRAS} is plotted with light gray crosses. The dashed horizontal line represents the injected power of simulated AGN jets $(5 \times 10^{45}~{\rm erg~s^{-1}})$.}
    \label{fig:LR_Ecav}
\end{figure*}

\begin{figure}
    \centering
    \includegraphics[width=\columnwidth]{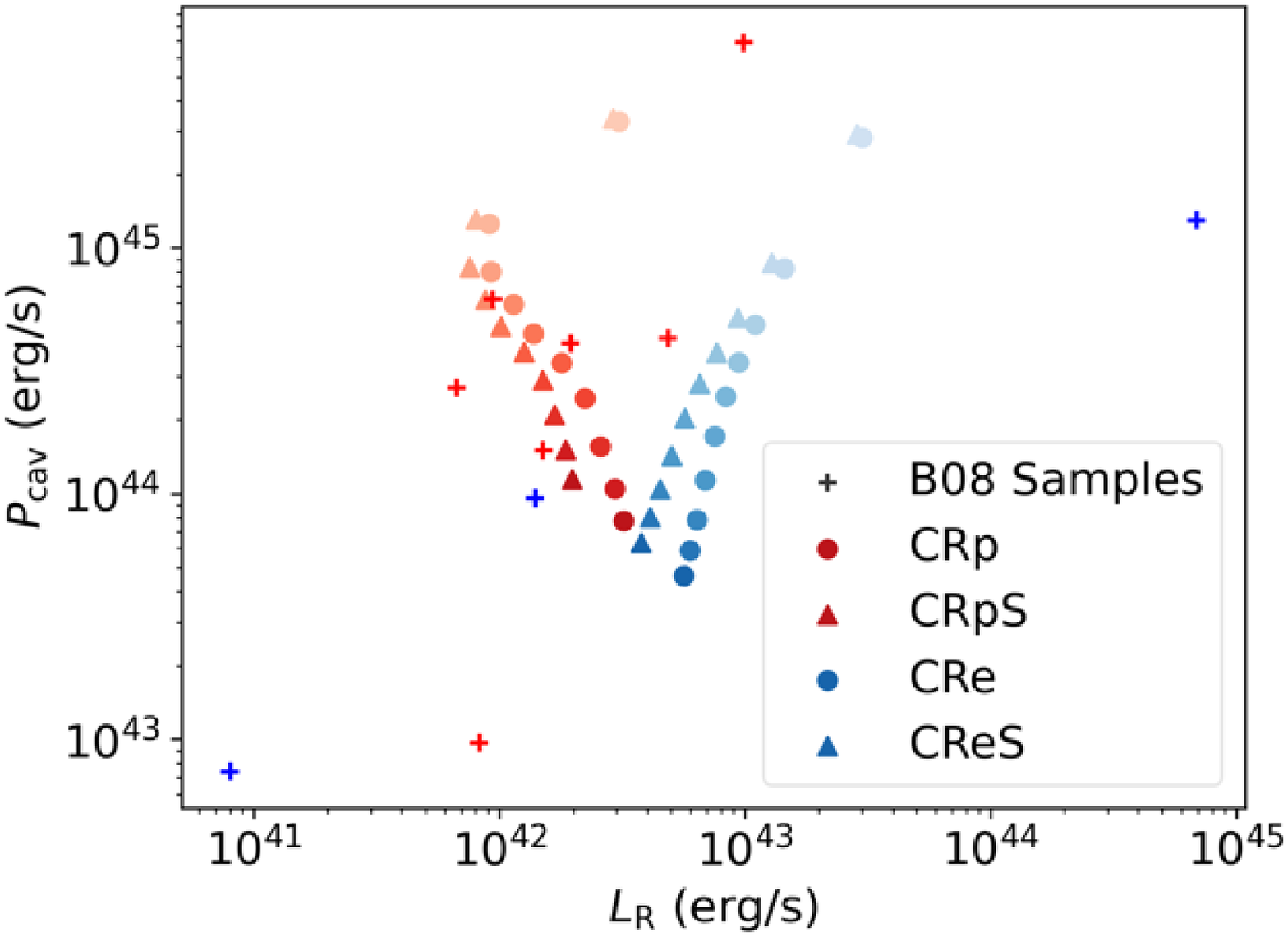}
    \caption{$P_{\mathrm{cav}}-L_\mathrm{R}$ diagram of our simulations over-plotted with observation data from \citet{Birzan08}. The simulation data points are plotted from $t=$ 5 to 95 Myr with a time interval of 10 Myr. Lighter colours represent earlier simulation times. We color-coded the observed data according to the $1+k$ values of the bubbles. The blue crosses represents the bubbles with $1+k$ < 100 (i.e., bubbles supported by radiating particles), while the red crosses represent bubbles with $1+k$ > 100 (i.e., bubbles supported by non-radiating particles).}
    \label{fig:LR_Ecav_B08}
\end{figure}

In Fig. \ref{fig:LR_Ecav} and Fig. \ref{fig:LR_Ecav_B08}, we calculate the observable properties of our simulated bubbles and compare them with observed cavities in \citet{Croston18}. This includes cavity samples from nearby CC clusters compiled by \citet{Birzan08}, which are mostly FRI sources, and a sample of other radio galaxies, taken from \citet{Cavagnolo2010ApJ,O'Sullivan2011ApJ, Ineson2017MNRAS}.
We compute the cavity power using $P_{\mathrm{cav}} = (E_{\rm CR}+E_{\rm th}+E_{\rm k}+PV)/t$, where $P$ is the total pressure inside the bubbles, $V$ is the volume of the bubbles, and $t$ is the simulation time, which is the age of the bubbles. This is compared with the observed cavity power, which is a proxy for the true jet power. Note that this definition is not strictly the same as that used by observers.\footnote{In general, the power of jets is not directly observable. For FRI jets, observers typically compute cavity power as $P_{\rm cav}=4PV/t_{\rm bub}$, where $t_{\rm bub}$ is the characteristic timescale (e.g. sound or buoyancy timescale) of the bubbles. For FRII jets, jet power can also be inferred from IC emission of the lobes in X-ray \citep[see e.g.][]{Ineson2017MNRAS}.} However, we argue that as observations advance, cavity power estimations should eventually approach to our definition of $P_{\rm cav}$. Therefore, our approach can still serve as a useful reference for future observations.
For the CRe/CReS simulations, the synchrotron luminosity is calculated using a pitch-angle averaged synchrotron emissivity of power-law CRe 
\begin{equation}
    \begin{split}
        \epsilon_{\mathrm{s}}(\nu)=\frac{e_{\mathrm{cr}}(2-p)}{E_{\mathrm{max}}^{2-p}-E_{\mathrm{min}}^{2-p}}(m_ec^2)^{1-p}\frac{3 \sigma_{\mathrm{T}} c  U_{\mathrm{B}}}{16 \pi \sqrt{\pi} \nu_{L}}\left(\frac{\nu}{\nu_{L}}\right)^{-\frac{p-1}{2}}\\
        \times 3^{\frac{p}{2}}\left( \frac{2.25}{p^{2.2}+0.105} \right)~ {\rm erg~cm^{-3}~s^{-1}},
    \label{eq: sync_epsilon}
    \end{split}
\end{equation}
\citep{Ghisellini2013}, 
where $\nu_L$ is the Larmor frequency. To 
obtain the total radio luminosity, we integrate over the whole simulation domain. For the CRp/CRpS cases, we consider that the synchrotron emission is dominated by the secondary particles created via hadronic collisions (mainly electrons and positrons from pion-decays). We model their spectrum following a steady-state approximation, where the secondary electron injection rate is balanced by their cooling (see \citealt{Owen2021} for details). We adopt a simplified analytic approximation for the production of secondary electrons, as detailed in Appendix~\ref{sec:secondary_electrons}.

From Fig.\ \ref{fig:LR_Ecav}, we observe the following features: 
\begin{enumerate}
    \item In general, CRe/CReS bubbles produce stronger synchrotron radiation than CRp/CRpS bubbles. This is because synchrotron emission from primary CRe is very efficient, whereas the production of secondary particles by the hadronic process occurs on relatively longer timescales.
    \item For simulations with CR streaming, the bubbles tend to produce slightly weaker radio emission. That is due to lower CR energy densities within the bubbles, resulting from streaming transferring energy from the CRs to the thermal gas.
    \item CRe/CReS bubbles reach their highest radio luminosity at early times ($< 10$) Myr. After the end of the injection phase, at 10 Myr, $E_{\mathrm{max}}$ and $E_{\mathrm{min}}$ quickly drop according to Eq.\ \ref{eq:E_lim_evo}, such that the radio emissivities (Eq.\ \ref{eq: sync_epsilon}) also gradually decrease as the bubbles age. 
    \item The radio luminosity of CRp/CRpS bubbles follows a different evolutionary path to CRe/CReS bubbles. Since synchrotron emission in the CRp/CRpS cases comes from secondary particles created by the hadronic process, the evolution of synchrotron luminosity is coupled with the hadronic heating rate. As described in \ref{sec:heating_coldgas}, rapid bubble expansion at early times dramatically reduces the gas density and the CR energy density near the cluster center, leading to a decrease in bubble synchrotron emission during the first 15 Myr. As the CRs diffuse out of the bubbles and interact with the ambient ICM at later times, the synchrotron emission rises with the corresponding increase of hadronic heating rates (see Fig.\ \ref{fig:heating_profile}).
    \item The predicted radio luminosities for both the CRe and CRp bubbles are broadly consistent with the observed ranges in the sample presented by \citet{Birzan08} (cf. Fig.\ \ref{fig:LR_Ecav}). 
\end{enumerate} 

The observational sample of bubbles from \cite{Birzan08} also provides information about their composition. This allows a more direct comparison to be made between our simulations and their data on the $P_{\rm cav}-L_{\rm R}$ plane. This is shown in Fig. \ref{fig:LR_Ecav_B08}, where the $x$-axis shows the total radio luminosity integrated from 10 to 10000 MHz. In \cite{Birzan08}, the composition of the bubbles was measured and presented in terms of $1+k$ values, where $k \equiv E_p/E_e$ is the energy ratio between non-radiating CRp and the radio-emitting CRe, i.e., $E_{\rm tot} = E_{\rm B} + (1+k)E_e$.
For bubbles that are not dominated by magnetic field energy, the $1+k$ value thus has the same physical meaning as the $P_{\rm ext}/P_{\rm int}$ value. In order to compare with the simulated CRp and CRe bubbles, we divide the sample in \cite{Birzan08} into two groups: one with $1+k>100$ (bubbles supported by non-radiating particles; shown in red crosses), and the other with $1+k<100$ (bubbles supported by radiating particles; shown in blue crosses). Note that Figs.\ \ref{fig:LR_Ecav} and \ref{fig:LR_Ecav_B08} are only different in the definition of the radio luminosity, and hence the trajectories of the simulated data points on the two plots are essentially identical.

Fig. \ref{fig:LR_Ecav_B08} shows that overall there is a good agreement between the simulated and observed data in terms of their locations on the $P_{\rm cav}-L_{\rm R}$ plane. There appears to be a tentative trend that the observed CRp bubbles have a larger slope on the $L_{\rm R} - P_\mathrm{cav}$ plane than the CRe bubbles. From the comparison with our simulated data points, the different slopes could potentially be explained by the different evolutionary tracks of the CRp and CRe bubbles.
However, with the current sample size, we were not able to draw a robust conclusion from this result. Future observational results to be obtained by the next-generation X-ray and radio facilities (e.g. {\it Athena} and the Square Kilometer Array, respectively) will allow fainter and more distant X-ray cavity systems to be explored. It would be interesting to consider in future work how bubbles with different $P_{\mathrm{int}}/P_{\mathrm{ext}}$ values would populate this diagram and provide insights into the evolution of bubbles of different compositions.

\section{Discussion}\label{sec:Discussion}

We find that the morphology, evolution, and the effect of ICM heating of CRp and CRe bubbles are very similar (see section~\ref{sec:Result}). This result is somewhat counter-intuitive, given that CRe provide no hadronic and negligible Coulomb heating in a bubble environment and undergo strong synchrotron and IC cooling, while CRp heat primarily through hadronic processes and do not experience significant synchrotron or IC cooling. Our findings can be understood as follows.

As shown in Fig. \ref{fig:E_EVO}, even though the CR energy within CRe bubbles drops rapidly, their total energy does not follow the same trend at later times. This is because the CRe bubbles become thermally-dominated by $t\sim 20$ Myr. After they become thermal bubbles, their total energies stop decreasing with the CR energy. Since the overall dynamical evolution of the bubbles depends on the total energy, the CRp and CRe bubbles follow similar dynamical evolution over long timescales. Therefore, the subsequent X-ray morphology of the CRp and CRe bubbles are also similar. Moreover, although synchrotron and IC cooling efficiently removes energy from the CRe, the bubbles can still heat up the ambient ICM efficiently via direct mixing \citep{Yang16b} because at later times the bubbles are thermally dominated. The amount of radio emission produced by CRp and CRe bubbles is also comparable -- in both cases the radio emission is within the observed ranges (see Fig. \ref{fig:LR_Ecav}). We therefore conclude that the evolution and feedback effects of AGN bubbles inflated by CRe dominated and CRp dominated jets are similar under the same initial conditions, and for the same initial jet power. It is difficult to determine the composition of an AGN bubble using its X-ray morphology or integrated radio luminosity alone.\footnote{The radio morphology of CRp and CRe bubbles could have systematic differences. As mentioned in Section \ref{sec:LRQjet}, synchrotron emission from CRp bubbles comes from the secondary particles that originate in hadronic collisions. Therefore, CRp bubbles tend to brighten at the bubble edges ($\propto e_{\rm cr}\rho^2$; cf. Eq. \ref{Eq:Hadronic_Cooling}). However, predicting realistic radio morphologies involves additional information (in particular, realistic CR spectral evolution, a more complete treatment of CR propagation and the coherence length of the initial tangled magnetic field). These details will be investigated in future, dedicated work.} 
It is even more difficult to infer the intrinsic composition of the jets by observing the bubbles they inflated because bubbles inflated by CRp and CRe jets would all become $P_{\rm ext}/P_{\rm int}\gg 1$ on longer timescales. Therefore, other means are required to distinguish their compositions.

One potential way to constrain the composition of AGN bubbles is by using the thermal Sunyaev--Zeldovich (SZ) effect \citep{SZ72, Birkinshaw99}. Because the thermal SZ effect directly traces the integrated thermal pressure along a line of sight, CR dominated bubbles would exhibit suppression in the SZ fluxes compared to the surrounding ICM, causing "SZ cavities" \citep{Pfrommer05, Ehlert18, Yang19}. Since our simulations show that bubbles created by CRp jets could remain CRp dominated for 70 Myr or more (Fig.\ \ref{fig:E_EVO}), and that bubbles created by CRe jets would become thermally dominated after $\sim 20$ Myr, the SZ effect could then be used to distinguish these two cases at later times of the bubble evolution.\footnote{Note, however, that the SZ effect would not be able to tell apart whether thermally dominated bubbles are created by CRe dominated jets or kinetic-energy dominated jets.}

One could potentially infer the composition of AGN jets/bubbles by tracking the evolution of the $P_{\rm ext}/P_{\rm int}$ value at early stages as the bubbles rise. One of the predictions from our simulations is that the $P_{\rm ext}/P_{\rm int}$ value for CRe bubbles keeps increasing as bubbles rise and lose CRe energy along the way. On the other hand, CRp bubbles have high $P_{\rm ext}/P_{\rm int}$ values throughout their evolution.\footnote{Since we define bubbles in the simulations using a cooling-time threshold, the $P_{\rm ext}/P_{\rm int}$ values are not affected by the mixing/contamination of the ambient thermal gas.} Therefore, by measuring the $P_{\rm ext}/P_{\rm int}$ values for a sample of AGN bubbles as a function of distance from the cluster center, one may be able to determine whether such an evolution exists and infer the intrinsic jet compositions. In fact, in the observed sample of AGN bubbles in the Perseus and Centaurus clusters \citep{Dunn2005MNRAS}, it is found that their $k$ values ($k\equiv E_p/E_e$) increase with the distance from the cluster center, implying that older and more distant bubbles tend to have greater pressure support from non-radiating particles. Particularly, the $k$ values for the bubbles in the Centaurus cluster increase from $\sim 1$ for the inner bubbles to $\sim 100$ for the outer bubbles (see their Fig.\ 8). Although different episodes of AGN jets do not necessarily have the same intrinsic composition, according to our simulations this observed trend would be more consistent with the evolution of AGN bubbles inflated by CRe dominated jets.       
Finally, we note that the observed trend of increasing $P_{\rm ext}/P_{\rm int}$ or $k$ values as a function of distance from cluster center suggests that there is no significant re-acceleration of CRe within the AGN bubbles \citep{Dunn2005MNRAS}. This conclusion is also supported by our simulations, which showed that the observed trends could be explained without invoking CR re-acceleration mechanisms. 

More recently, \citet{Vazza2021,Vazza2022arXiv} have performed detailed analysis on the transport and energetics of relativistic electrons using cosmological simulations. They found that even when both shock (Fermi I) and turbulence (Fermi II) re-acceleration mechanisms are considered, the energy density of CRe would not exceed $\sim$ 10 per cent of the local thermal gas energy \citep[see e.g. Fig. 12 and 13 in][]{Vazza2022arXiv}). Their results therefore also support that the re-accelerated CRe represent a sub-dominant component in the dynamics of AGN bubbles. If this is the case, tracing the relation between $P_{\rm ext}/P_{\rm int}$ and the distance of the bubbles from cluster center within the same cluster can in principle provide important clues on the intrinsic composition of AGN jets.

\section{Conclusions}\label{sec:Conclusion}
Relativistic jets from SMBHs are one of the most important yet complicated mechanisms that could affect galaxy evolution and provide energetic feedback to the ICM in CC clusters. However, the feedback effects to the ICM by AGN jets/bubbles with different energy compositions remain poorly understood. Observational constraints of cluster radio bubbles suggest that there could be two populations of AGN bubbles: one dominated by radiating particles (i.e., CRe), the other dominated by non-radiating particles (e.g., CRp). In order to understand the evolution of AGN bubbles and their influence on the AGN feedback mechanisms in these two scenarios, we performed four 3D MHD simulations of CRp-dominated jets versus CRe-dominated jets, with and without CR streaming (Table \ref{tab:Sim_Type}). We investigated their differences in terms of the dynamical evolution, the amount of heating provided to the ICM, and their observable properties in the X-ray and radio bands. We summarize the key results as follows.
\begin{enumerate}
    \item Despite the stronger synchrotron and IC cooling of CRe, the long-term evolution of bubbles inflated by CRe jets is very similar to those by CRp jets (Fig. \ref{fig:density_evo}). This is because, although the energy of CRe within the bubbles is quickly lost due to synchrotron and IC cooling, thermal energy takes over and becomes the dominant energy component within $\sim 20$ Myr (Fig. \ref{fig:E_EVO}). Afterwards, the total bubble energy stops decreasing with the rapidly declining CR energy, and hence the bubbles have similar dynamical evolution to the CRp dominated bubbles.
    \item All four simulations in our study show very similar amount of cold gas formed via local thermal instabilities (Fig. \ref{fig:coldmass5e5}), suggesting that the ability of CRp and CRe bubbles to heat the ICM is similar. In addition, the CR heating rates in all four simulations are much weaker than the radiative cooling rate (except for CRp cases at earliest times), even though all simulations (including a case with no heating from CRs, i.e., the CRe simulation in Table \ref{tab:Sim_Type}) show suppressed formation of cold gas at a similar level. These results suggest that, in addition to heating from CRs (Coulomb, hadronic, and streaming), other heating mechanisms such as direct mixing still play an important role.
    \item With CR streaming, the bubbles can provide stronger heating to the ICM and the heating can extend to larger radii than cases without streaming (Fig. \ref{fig:heating_profile}). The heating rates of CRe bubbles are in general smaller than the CRp bubbles because of the lower CR energy density due to cooling. For both the CRp and CRe cases, CR streaming could help to reduce the amount of cold gas within the simulations by $\sim 20\%$. 
    \item We computed the predicted radio luminosity for the CRp and CRe bubbles and investigated their evolution on the $P_\mathrm{cav}-L_\mathrm{151~MHz}$ plane. 
    We find that the CRp and CRe bubbles evolve differently because of their different emission mechanisms. For CRe bubbles, their synchrotron emission decreases with time due to the energy losses of CRe. For CRp bubbles, their synchrotron emission comes from secondary electrons produced by hadronic interaction processes. This is suppressed at early times, when rapid expansion of the bubbles reduces the gas density close to the cluster center. 
    Despite the difference in their evolutionary trajectories, the predicted radio luminosity for both the CRp and CRe bubbles in our simulations are broadly consistent with the observed FRI sample from \citep{Birzan08} (Fig. \ref{fig:LR_Ecav}).
    \item Overall, we find that AGN bubbles inflated by CRe dominated and CRp dominated jets behave very similarly in terms of their dynamical evolution, X-ray morphology, their ability to heat the ICM and suppress cold-gas formation, as well as their radio luminosity.
    Our results suggest that it may be difficult to determine the composition of an AGN bubble using its X-ray morphology or integrated radio luminosity alone, and inferring the intrinsic jet composition from the these bubbles may be even more challenging. Other observational techniques (e.g., the SZ effect) would be needed to help constrain their composition.  
    \item Our simulations predict that, due to the cooling of CRe, the $P_{\rm ext}/P_{\rm int}$ values (or, equivalently, the ratio between non-radiating and radiating particles $k\equiv E_p/E_e$ in the literature) for CRe bubbles would increase as the bubbles rise toward larger radii. In contrast, the CRp bubbles would have high $P_{\rm ext}/P_{\rm int}$ values throughout their evolution. Therefore, measuring the $P_{\rm ext}/P_{\rm int}$ values as a function of distance from cluster centers for bubbles within the same cluster could potentially provide additional constraints on the composition of AGN jets/bubbles. Interestingly, the cavities in the Centaurus cluster \citep{Dunn2005MNRAS} show increasing $k$ values with distance from the cluster center, suggesting that these bubbles could be produced by CRe dominated jets. In addition, the fact that both the simulated and observed bubbles show rising $k$ values with distance suggests that re-acceleration of CRe is subdominant within AGN bubbles.
\end{enumerate}

\section*{Acknowledgements}
YHL and HYKY acknowledge support from National Science and Technology Council (NSTC) of Taiwan (109-2112-M-007-037-MY3). HYKY acknowledges support from Yushan Scholar Program of the Ministry of Education (MoE) of Taiwan. ERO is an overseas researcher under the Postdoctoral Fellowship of Japan Society for the Promotion
of Science (JSPS), supported by JSPS KAKENHI Grant Number JP22F22327, 
and also acknowledges support from the Center for Informatics and Computation in Astronomy (CICA) at National Tsing Hua University (NTHU) through a grant from the MoE of Taiwan. This work used high-performance computing facilities operated by CICA at NTHU. FLASH was developed largely by the DOE-supported ASC/Alliances Center for Astrophysical Thermonuclear Flashes at University of Chicago. Data analysis presented in this paper was conducted with the publicly available yt visualization software \citep{yt}. We are grateful to the yt development team and community for their support. This research has made use of NASA's Astrophysics Data Systems. We thank the anonymous referee for their insightful comments and suggestions that led to improvements in our manuscript.

%%%%%%%%%%%%%%%%%%%%%%%%%%%%%%%%%%%%%%%%%%%%%%%%%%
\section*{Data Availability}
The data underlying this article will be shared on reasonable request
to the corresponding authors.



\bibliographystyle{mnras}
\bibliography{agn} % if your bibtex file is called example.bib

%%%%%%%%%%%%%%%%%%%%%%%%%%%%%%%%%%%%%%%%%%%%%%%%%%

%%%%%%%%%%%%%%%%% APPENDICES %%%%%%%%%%%%%%%%%%%%%

\appendix

\section{Influence of bubble definition}\label{app: BubbleDef}
\begin{figure*}
    \centering
    \includegraphics[width=\textwidth]{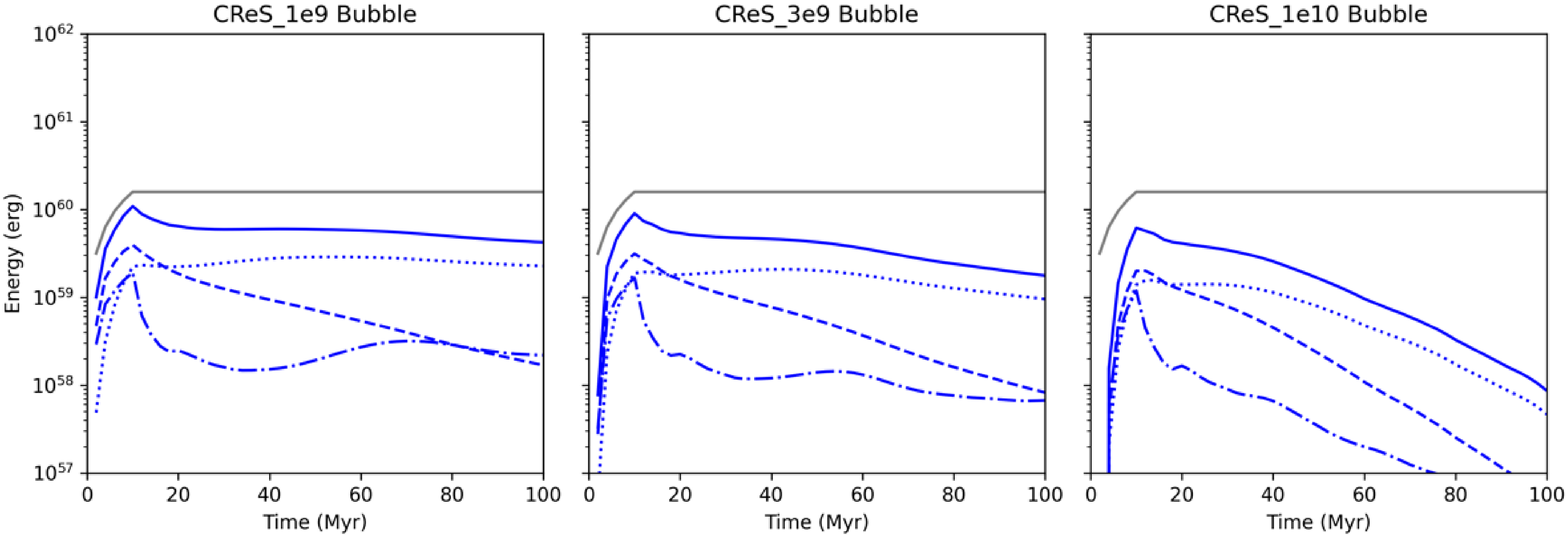}
    \caption{Convergence test on the choice of Bubble definition in CReS simulation. In each Panel, we show the energy evolution inside the bubbles calculated using different cooling time criteria (1 Gyr, 3 Gyr and 10 Gyr from left to right).}
    \label{fig:BubbleDef_Convergence}
\end{figure*}

Here, we assess whether our simulation results would be influenced by the definition of the bubbles - in particular, their energy evolution. Fig. \ref{fig:BubbleDef_Convergence} shows the evolution of bubble energy in the CReS simulation. From left to right, we define the bubbles using radiative cooling times of 1, 3 and 10 Gyr, respectively. Despite the apparent difference in their exact values, the main features (relative strength of different energy components and crossover time between the thermal and CR energies), are preserved regardless of the definition we adopt. We conclude our main results are not sensitive to the definition of bubbles in the simulations.

\section{Influence of the choices of maximum and minimum electron energy}\label{sec:EmaxEmin}
\begin{figure*}
    \centering
    \includegraphics[width=0.8\textwidth]{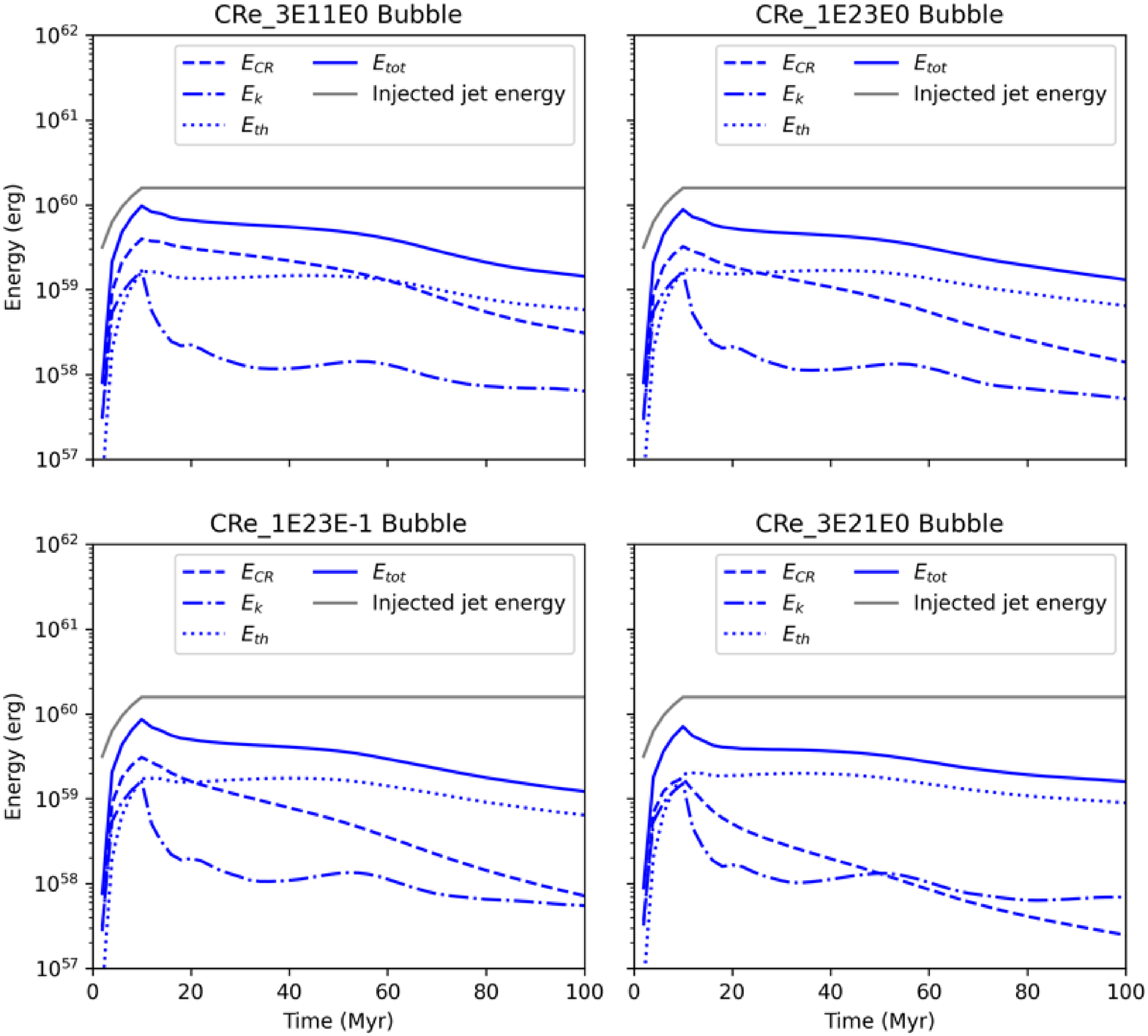}
    \caption{Convergence test on the choice of $E_\mathrm{min,0}$ and $E_\mathrm{max,0}$. The titles of the plots represent their energy range. For example, CRe\_3E11E0 represent $E_\mathrm{max,0}=30~\mathrm{GeV}$, $E_\mathrm{min,0}=1~\mathrm{GeV}$}
    \label{fig:EminEmax_Convergence}
\end{figure*}

In this section, we demonstrate how the choices of $E_\mathrm{min,0}$ and $E_\mathrm{max,0}$ would affect the energy evolution of the simulated AGN bubbles. We performed four CReS simulations with $E_\mathrm{max,0}$ and $E_\mathrm{min,0}$ lies between 300 GeV to 300 MeV. We show their energy evolution in Fig. \ref{fig:EminEmax_Convergence}. We find that, in general, higher $E_\mathrm{max,0}$ and $E_\mathrm{min,0}$ would result in stronger cooling, bringing the crossover between thermal and CR energies earlier in time. For simulations with $(E_\mathrm{max,0}, E_\mathrm{min,0})=(100~\mathrm{GeV}, 3~\mathrm{GeV})$ and $(100~\mathrm{GeV}, 0.3~\mathrm{GeV})$, the results do not change significantly compared to the fiducial setup. For the simulation with $(E_\mathrm{max,0}, E_\mathrm{min,0})=(300~\mathrm{GeV}, 1~\mathrm{GeV})$, the CR energy becomes sub-dominant right after the end of jet injection due to strong cooling of high energy CRe. For the simulation with $(E_\mathrm{max,0}, E_\mathrm{min,0})=(30~\mathrm{GeV}, 1~\mathrm{GeV})$, the crossover between thermal and CR energies is delayed to roughly 60 Myr since low-energy CRe have longer cooling times. 

Our tests suggest that, even for the conservative case where the initial energy of CRe is low, the bubbles cannot remain CRe supported after $\sim 60$ Myr. Therefore, if future observations find CRe bubbles with similar properties to our simulation setup and with age greater than 60 Myr, they are highly unlikely to be supported by primary CRe. 
We emphasize that we do not aim to directly compare our results to observed bubbles, since their age estimation is often highly uncertain. 
Instead, we focus on demonstrating of the importance of the underlying physical mechanisms. Overall, our main conclusions hold regardless of the choices of $E_\mathrm{max,0}$ and $E_\mathrm{min,0}$.

\section{Secondary CR production in hadronic collisions}
\label{sec:secondary_electrons}

To compute 
the synchrotron luminosity of our CRp and CRpS simulations, 
we require the steady-state spectrum of the secondary electrons (hereafter we refer to both electrons and positrons produced in hadronic collisions as `electrons'). This is obtained by balancing the injection of the electrons against their cooling. The secondary electron injection rate in hadronic pp collisions is given by \citep{Owen2021}: 
\begin{equation}
Q_{\rm e}(\gamma_{\rm e}) =  \int_{\gamma_{\rm 0, p}}^{\gamma_{\rm 1, p}}\frac{\partial f(\gamma_{\rm e};\gamma_{\rm p})}{\partial \gamma_{\rm p}} \;\! \dot{n}_{\rm p\pi}(\gamma_{\rm p}) \;\! {\rm d}\gamma_{\rm p} \ ,
\label{eq:injection_secondaries}
\end{equation}
where the energetics of the pp interaction are specified by the Lorentz factor of the initiating CR proton, $\gamma_{\rm p}$, and where $\gamma_{\rm 1,p}$ and $\gamma_{\rm 0, p}$ are set by the maximum and minimum limits of the CRp spectrum, respectively. Here $\gamma_{\rm e}$ is the secondary CRe Lorentz factor, $\dot{n}_{\rm p\pi}(\gamma_{\rm p})$ is the weakly energy-dependent pp interaction cross section~\citep[for which we adopt the parametrization obtained by][]{Kafexhiu2014}, and ${\partial f(\gamma_{\rm e};\gamma_{\rm p})}/{\partial \gamma_{\rm p}}$ is relative production fraction of electrons in the proton's rest frame.

In the hadronic pp interaction channel, the production of charged pions mediates the formation of muons, $\mu^{\pm}$, for which it is assumed that roughly each charged pion will decay into a single muon. The Lorentz factors of the muons that form in the decay can be related to $\gamma_{\rm p}$ by 
\begin{equation}
\label{eq:p_mu_gammarel}
\gamma_{\rm p} = \frac{1}{{\kappa}_{\rm p \pi}} \;\! \left[ 8 \gamma_{\mu}^2 \bar{m}^2 - 1\right],
\end{equation}
where we adopt the notation for dimensionless combined rest mass as
\begin{equation}
    \bar{m} = \frac{m_{\rm \pi}\;\!m_{\rm \mu}}{m_{\rm \pi}^2 + m_{\rm \mu}^2} \ ,
\end{equation}
with $m_{\rm \pi}$ and $m_{\rm \mu}$ as pion and muon rest masses, respectively. $\kappa_{\rm p\pi}(\gamma_{\rm p}) \approx 2\;\!f_{\nu}/[3\mathcal{M}_{\rm p\pi}(\gamma_{\rm p})]$ quantifies the average fraction of pion energy that is passed to the muons via pion production, with $f_{\nu} = 3/4$ accounting for the fractional energy loss to neutrinos (assumed 25\% -- e.g.~\citealt{Burbidge1956PhRv, Loeb2006JCAP, Lacki2013MNRAS}). $\mathcal{M}_{\rm p\pi}(\gamma_{\rm p})$ is introduced as the energy-dependent charged pp pion multiplicity (cf.~\citealt{Owen2019AA}, which adopted the parameterisation of~\citealt{Albini1976NCimA} and~\citealt{Fiete2010JPhG}). Strictly, this is the channel multiplicity, so reflects the multiplicity of charged products. Since a single charged pion will yield one charged muon which, in turn, produces one charged electron, the channel multiplicity can be equivalently taken as the charged pion, muon or electron multiplicity as there is equivalence with the number of channels.

Electrons are formed from the decay of the muons. The distribution of their energies in the muon rest frame is given by
\begin{equation}
    \frac{{\rm d}N(\gamma_{\rm e}^*)}{{\rm d}\gamma_{\rm e}^*} = \frac{8 (\gamma_{\rm e}^*)^2 \;\! m_{\rm e}^3}{m_{\rm \mu}^3}\left(3-4\frac{\gamma_{\rm e}^*\;\!m_{\rm e}}{m_{\rm \mu}}\right) \ ,
\end{equation}
if their emission in the rest frame of the parent muon is assumed to be isotropic~\citep{Berrington2003ApJ, Dermer2009_book}. In the lab frame, this transforms to (using $\gamma_{\rm e} = \gamma_{\rm \mu}\gamma_{\rm e}^*$ for $\gamma_{\rm e}^*$ as the Lorentz factor of the electrons in the muon rest frame)
\begin{equation}
    \frac{{\partial}N(\gamma_{\rm e}; \gamma_{\rm \mu})}{{\partial}\gamma_{\rm \mu}} = \frac{8}{\gamma_{\rm \mu}}\left(\frac{\gamma_{\rm e}m_{\rm e}}{\gamma_{\rm \mu}m_{\rm \mu}}\right)^3\;\!\left(4\left[\frac{\gamma_{\rm e}m_{\rm e}}{\gamma_{\rm \mu}m_{\rm \mu}}\right] - 3 \right) \ ,
\end{equation}
such that a relative production fraction of electrons of energy $\gamma_{\rm e}$ can be defined as
\begin{equation}
\label{eq:prod_frac_mu_elec}
   \frac{\partial f(\gamma_{\rm e};\gamma_{\rm \mu})}{\partial \gamma_{\rm \mu}} = \frac{4 \gamma_{\rm e} m_{\rm e} - 3 \gamma_{\rm \mu} m_{\rm \mu}}{\gamma_{\rm \mu}^5(\gamma_{\rm e}m_{\rm e} - m_{\rm \mu})} \ ,
\end{equation}
which quantifies the relative production rate of electrons with energy $\gamma_{\rm e}$ compared to the total production rate of all electrons due to the decay of muons of energy $\gamma_{\rm \mu}$.
The overall relative production fraction of electrons in the initiating proton's rest frame for the full pp interaction chain then follows by substitution of equation~\ref{eq:p_mu_gammarel} into equation~\ref{eq:prod_frac_mu_elec}, to give:
\begin{equation}
    \frac{\partial f(\gamma_{\rm e};\gamma_{\rm p})}{\partial \gamma_{\rm p}} = \frac{4 \gamma_{\rm e} m_{\rm e} \bar{m}^5 \kappa_{\rm p \pi}  - 3 m_{\rm \mu} \bar{m}^4 {\kappa}_{\rm p \pi} \left( \gamma_{\rm p} {\kappa}_{\rm p \pi} + 1 \right)^{1/2}}{2 \left( \gamma_{\rm p} {\kappa}_{\rm p \pi} + 1 \right)^{3} (\gamma_{\rm e}m_{\rm e} - m_{\rm \mu})} \ .
    \label{eq:final_elec_effic_pp}
\end{equation}
This is used in calculating the produced spectrum of secondary electrons from the pp interaction.

%%%%%%%%%%%%%%%%%%%%%%%%%%%%%%%%%%%%%%%%%%%%%%%%%%

% Don't change these lines
\bsp	% typesetting comment
\label{lastpage}
\end{document}